\documentclass[twocolumn,aps,prl,groupedaddress,amssymb]{revtex4}
\usepackage{graphicx}
\usepackage{dcolumn}
\usepackage{amsmath}
\begin{document}

\title{  Evanescent waves in photonic crystals 
 and image of Veselago lens }

\author{A. L. Efros}
\email{efros@physics.utah.edu}
\affiliation{University of Utah, Salt Lake City UT, 84112 USA}

\author{C. Y. Li}
\affiliation{University of Utah, Salt Lake City UT, 84112 USA}

\author{A. L. Pokrovsky}
\affiliation{University of Utah, Salt Lake City UT, 84112 USA}

\date{\today}

\begin{abstract}
It is shown that negative electric permittivity $\epsilon$ and magnetic permeability $\mu$ recently discovered in a photonic crystal in the vicinity of the 
$\Gamma$-point are properties of propagating modes only. The evanescent modes rather decay than increase in the bulk of the crystal though they may be
 amplified by  surface waves. If  surface support such waves, the evanescent waves may  improve   the image of a thin  Veselago lens. It is shown that  a ``perfect lens'' contradicts to the wave optics and  a criterion of ``superlensing'' is formulated.
\end{abstract}
\maketitle

 Left handed medium (LHM), defined by Veselago\cite{ves} as a medium with simultaneously negative $\mu$ and $\epsilon$, recently attracted  much attention mostly because of the negative refraction at the interface with a regular medium (RM). This effect allows creation of a unique device called the ``Veselago lens''. This lens is a slab of  LHM inside a RM with a condition that  both media have the same {\em isotropic} refractive index and the same impedance. The interest in  LHM was significantly amplified by the work of 
Pendry\cite{pen} who argued that 
the Veselago lens is a ``perfect lens'' in the sense that it gives a perfect  image of the point source.
  This statement is based upon the observation that the evanescent waves (EW's) of a form $\exp (ik_y y -\kappa x)$ that usually decay in the near field region are amplified by the LHM. Pendry claimed that the amplified EW's  may restore a perfect image. Following Veselago, Pendry
considered  a "hypothetical" LHM (HLHM) with negative and real $\mu$ and $\epsilon$. He did not present a solution in a coordinate space. This was done by Ziolkowski and Heyman\cite{ziol} and their solution reveals a fundamental problem with  Pendry's idea.    The solution diverges exponentially at each point of a 3D domain near the focus, just where fields of EW's increase due to  amplification.  This was found out almost simultaneously by three groups\cite{gar,pok3,hal}.
 The first group argued that the EW's  should just be omitted as non-physical. 
The second group used diffraction theory,  known to be exact at small wave length, and found a finite width of the focus. Haldane\cite{hal} argued  that the problem does not have a solution in the framework of macroscopic
 electrodynamics.

 We consider here a dielectric  photonic 
crystal (PC)  that is known to be a real LHM in some frequency range\cite{ef, pok}. 
 We show   that its electrodynamics  differs from that of  the HLHM and that
in the real system negative  $\mu$ and $\epsilon$ of propagating waves do not provide 
   amplification of EW's. We present also a general proof that the perfect imaging of a point source contradicts  wave optics independent of the nature of a
 lens.

 A  large  group of works pioneered by 
Luo {\it et al.}\cite{joan,joan2} claims observation of  negative refraction, focusing, and even 
superlensing  in a PC due to   photons with wave vectors deep in the
 Brillouin zone.
Propagation of such photons can be described macroscopically only taking into account spatial dispersion (${\bf k}$-dependence of $\epsilon$). 
The theory of Veselago can hardly be generalized for the case of  spatial dispersion mostly because of an extra term in the definition of Poynting's vector\cite{agr}. Following Ref.\cite{joan} we do  not consider such a medium left handed and do not discuss it here.

It was shown recently\cite{ef, pok} that a 2D dielectric uniaxial PC made of non-magnetic
 materials 
can behave as a LHM with negative $\epsilon$ and $\mu$ if it has a negative group velocity 
in the vicinity of the $\Gamma$-point. This was proved for propagating modes only.
Experimental demonstration of negative refraction in a metallic PC using the modes near the $\Gamma$-point has been presented in Ref. \cite{parimis}.

\begin{figure}
\includegraphics[width=8.6cm]{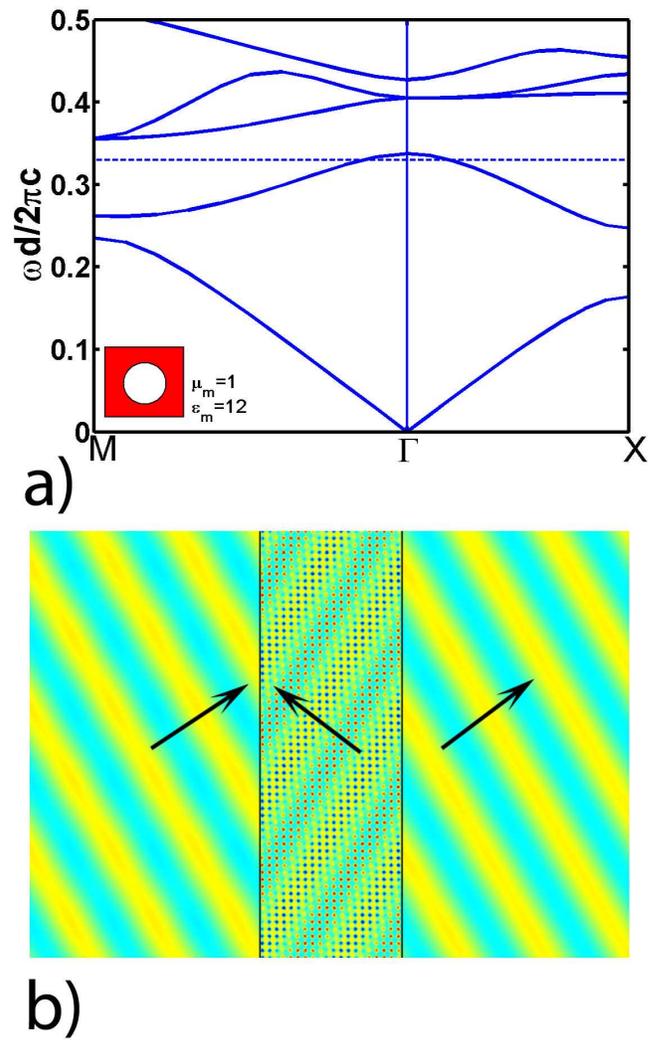}
\caption{ (Color online) a) Five lowest bands of the photonic spectrum of a 2D PC with a period $d$.
The unit cell of the PC is shown in the inset. The PC consists of 
a square lattice of circular cylindrical holes in a dielectric matrix with $\epsilon_m=12$, $\mu_m=1$.  
The radii of the holes $R=0.35d$.
b) Negative refraction of 
 light propagating through the PC slab surrounding by homogeneous medium  at the working frequency. Periodic boundary conditions are used in the vertical direction.
The arrows show the directions of the wave vector.
 \label{fig1}}
\end{figure}

In this paper we consider p-polarization for both EW's and propagating modes in a  uniaxial PC. Fig. \ref{fig1}(a) shows the  elementary cell in the plane 
$x-y$ and  the spectrum of   p-polarized propagating waves. 
 The dashed line is our working frequency. All data below are obtained for this frequency.

 From the same numerical solution of microscopic Maxwell's equation that gives
 the  spectrum  one gets microscopic fields $h_z(x,y)$, $e_y(x,y)$ and $e_{x}(x,y)$ corresponding to a given frequency of propagating modes. The  microscopic fields are the Bloch functions wile  the macroscopic fields 
 have a form  $B_z=<$$h_z$$>$, ${\bf E}=$$<$${\bf e}$$>$, where  $<$...$>$ means averaging over the unit cell. Near the $\Gamma$-point the Bloch functions have a small ${\bf k}$, and 
the macroscopic fields of the propagating modes are plane waves that obey the 
macroscopic Maxwell equations.

As follows from Ref.\cite{pok}, the values of $ \mu_{zz}\equiv \mu$  and $\epsilon_{xx}=\epsilon_{yy}\equiv \epsilon$,  that describe propagation of these waves,  are negative because their  group velocity is negative.
 Here $z$ is the axis of the crystal. It has been shown previously\cite{ef} that the values of $\epsilon$ and $\mu$ can be expressed through the integrals of 
Bloch's function at $k=0$. It is important that in a system with a spatial dispersion the separation of magnetization and displacement currents is impossible
\cite{lan}.  Our parameters  $\epsilon(\omega)$ and $\mu(\omega)$ are not  exclusive  properties of the medium at a given frequency, as  in  macroscopic electrodynamics without spatial dispersion. Rather they  are   properties of a particular mode of this medium, whose fields are used for the  calculations. We argue 
below that they are not applicable to the EW's. The theorem\cite{pok} connecting signs of  both
 $\epsilon(\omega)$ and $\mu(\omega)$ with the sign of group velocity is not applicable to the EW's as well.   Near the $\Gamma$-point 
$\mu\sim k^2$\cite{ef}. Since the dispersion law $\omega(k)$ is also isotropic one obtains $\mu$ that are function of $\omega$ only. This approach is good for an infinite PC, but may create problems with boundary conditions\cite{agr} because in coordinate space $k^2$
should be considered as Laplacian operator. We are going to refer to this problem somewhere else. For the EW's this approach fails completely because their 
dispersion law is anisotropic.

To demonstrate Veselago lens one should know  $\mu$ and $\epsilon$ of the RM 
medium around the PC.  In this paper we find  the value of of refractive index $n=\sqrt{\epsilon\mu}$ from the function  $\omega(k)$, shown in Fig. \ref{fig1}(a), using definition $\omega=kc/n$. The second condition that provides matching of impedance  is the absence of 
reflection of the incident plane wave  in a wide range of incident angles.

Using these two conditions we get that at the working frequency the RM material should have $\epsilon = 1.125, \mu = 0.08$.
The physical reasons for the appearance of $\mu$ in a non-magnetic PC   are discussed in Ref. \cite{ef}. In all computations below we use these values for 
 homogeneous regions surrounding the PC slab.

Fig. \ref{fig1}(b) shows the result of our simulation for  light propagation through the PC slab surrounded
by a homogeneous medium.
One can see the negative refraction of  light  coming in and going out of the PC with equal absolute values of incident and refracted angles and without any
visible  reflections  from the interfaces. We checked it in a wide range of the incident angles (from 0 to 60 degrees). An animation\cite{web} shows that the wave fronts are moving to the right outside  the PC slab and to the left inside the slab. All of these phenomena exactly correspond to Veselago's theory\cite{ves}.

\begin{figure}
\includegraphics[width=8.6cm]{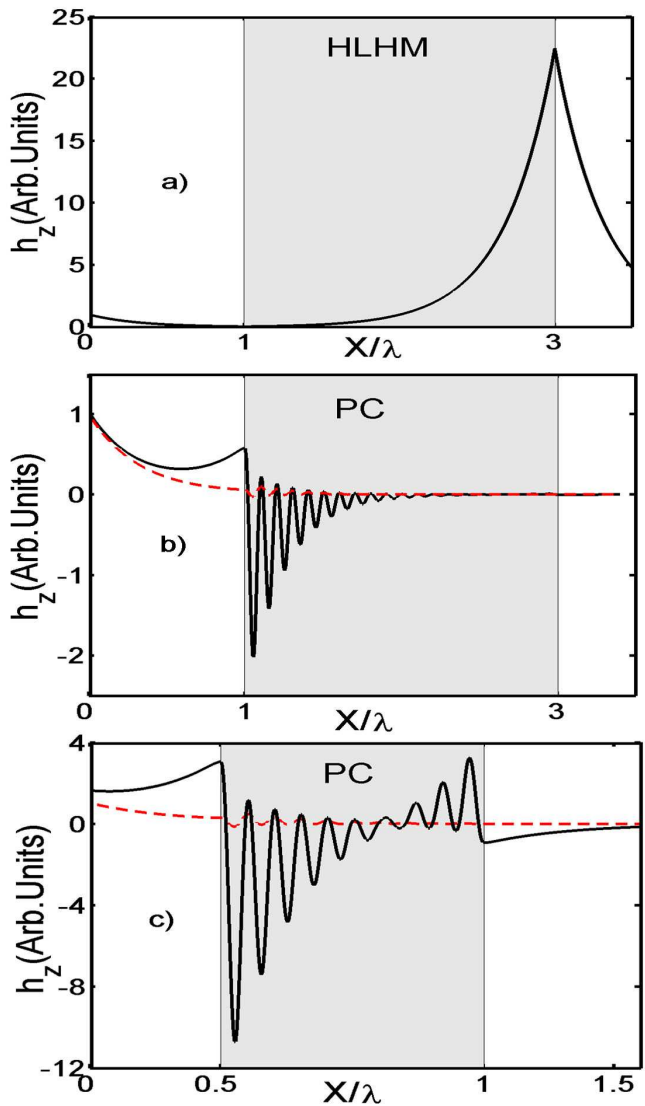}
\caption{(Color online) Magnetic field of the EW in the HLHM (a) and in the PC slabs (b),(c) surrounded by a regular homogeneous medium at the following parameters of EW: $k_y = (\sqrt{5}/2) k_0$, $\kappa=0.5k_0$, $k_0=\omega n/c$, $n=\sqrt{\epsilon \mu}=0.3$.  The cross section $y=0$ is shown. The thickness of the PC
 slab is 20$d$ (b) and 10$d$ (c). The full lines (black) are for the AH
 surfaces,
the dashed  lines (red) are for BH surfaces. The period of oscillation is $d$. 
 \label{fig2}}
\end{figure}

Now we come to EW's in the same PC slab imbedded into the same RM as in Fig 1 (b) . Our results are shown in Fig. 2. The boundary conditions implied for an incident wave at  $x=0$
 are  $h_z = \exp{(i k_y y)}$ and  $d h_z / dx = -\kappa h_z$. Note that due to reflection the total field shown in Fig. 2 may  not coincide at $x=0$ with the incident one. To obey Maxwell's equation we imply the dispersion law of EW's in a form $k_y^2-\kappa^2=\omega^2 n^2/c^2$. The distance from the left boundary of the RM (point $x=0$) to the PC slab is 1/2 of the slab thickness. In Fig. 2 (a)the PC slab is substituted by the HLHM that has negative $\epsilon$ and $\mu$ with the  absolute values
same as in the RM.  One can see a strong amplification of the EW first discovered by Pendry\cite{pen}. However the PC slab gives a very different picture. The result strongly depend on the way the surface is cut. Our surfaces are perpendicular to [10] direction, but they cut either across the holes (AH) or between the holes (BH). In the case of the  BH surface (dashed lines) the EW decreases in the PC slab in the same way as in RM without any signature of  amplification. However, there is an amplification at  the left AH surface for a thick slab and at both AH surfaces for a thin slab. We think that this amplification are due to the surface waves (SW), excited by EW's. Since  the picture is not symmetric with respect to left and right interfaces this excitation is not resonant.  These SW's have nothing in common with left-handed properties and with polarons in the HLHM, considered by Ruppin\cite{rup} because they depend on the properties of the surface. We think these SW's appear just due  to the termination of periodic structure. They have been studied before by many authors\cite{jo,joan2,ha,souk} with and without connection to the focusing.

\begin{figure}
\includegraphics[width=8.6cm]{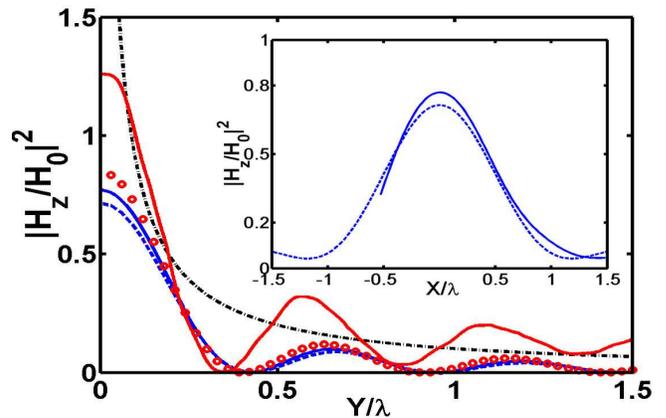}
\caption{ (Color online) Distribution of magnetic energy near the focus of the lens in lateral and perpendicular directions. Solid lines show results of simulation. The upper curve (red) is distribution in lateral direction for a thin lens 
  with AH surfaces, the lower curve(blue) is the same for a thick lens. Open circles
 (red)
 shows the analytical result for a thin lens, the dashed curve (blue) shows the same for a thick lens.   Dash-dotted line (black) shows the  distribution of magnetic energy of 
   the point source.  The inset shows computational  distribution in perpendicular direction(solid blue)
 and 
 analytical result in this direction (dashed blue) both for the  thick lens only. 
 \label{fig3}}
\end{figure}

 Thus, we have found that  EW's
cannot be described in terms of the same  negative    $\epsilon$ and $\mu$ that are  found for the 
propagating waves. The formal reason is that the  dispersion law of EW is anisotropic even near the $\Gamma$-point. Therefore in this case $\mu$ cannot be represented in ${\bf k}$-independent form as in the case of propagating waves. The physical explanation reads that the EW in the direction of decay is the same for both  regular and backward wave. Therefore in this direction there is no difference  between LHM and RM.  We have verified, however, that if 
an  EW propagates (has a real $k$) in a direction not parallel to the surface of the PC slab,
 it shows a negative refraction, similar to that shown in Fig. 1(b).

Now we describe the image of the Veselago lens.  The point source
 is the 2D Green function $H_s=iH_0H_0^{(1)}(\rho k_0)\exp{-i\omega t}$,
 where $k_0=
\omega n/c$ and 
the Hankel function $H^{(1)}_0=J_0+iN_0$. It is located in the RM at a distance $a$ to  the left of the PC slab.

  It is easy to show that   that the perfect image of this source is impossible. Indeed, if the image has the same form as the source,
 the magnetic field near the focus would obey inhomogeneous Helmholtz equation with a delta-function at the focus, while the  focal point would  not contain any source. Thus, this  solution does not obey homogeneous  Helmholtz equation near the focal point. Note that this argument does not specify the nature of a lens.

We simulated the Veselago lens with the same PC slab and the same parameters 
of the RM.   Similar to geometry of Fig. 2 the distance $a$ is   1/2 of the thickness of the PC slab. Fig. 3 shows the spatial  distributions of magnetic energy near the focus in both lateral ($y$) and perpendicular ($x$) directions. The distribution in the lateral direction is symmetric for $y$ and $-y$ so it is  shown for positive $y$ only.

Our results shown in Fig. 2  give   a possibility to calculate analytically the distribution of energy  near  the focus of a thick Veselago lens with both AH and BH surfaces or of a  thin lens with BH surface. In these cases evanescent waves do not reach the focus.  At positive $x$ the Green function of the source  can be represented in a form $H_s=H_p+
H_{ev}$, where 
\begin{equation}
H_p=i(H_0/\pi)\int_{-k_0}^{k_0} {\exp{i\left(k y+x\sqrt{k_0^2-k^2}-\omega t\right)}\over
\sqrt{k_0^2-k^2}}dk 
\label{p}
\end{equation}
contains only propagating modes while 
\begin{equation}
H_{ev}=(H_0/\pi)\int_{|k|>k_0} \frac{\exp\left(ik y-x\sqrt{k^2-k_0^2}-i\omega t\right)}{\sqrt{k^2-k_0^2}}dk 
\label{ev}
\end{equation}
contains only EW's.

If the  EW's decay inside the PC slab, only term  $H_p$ contributes to the focus if the slab is thick enough . On the other hand our results show that for the propagating modes the PC slab works exactly like a HLHM. Using the Fresnel equation it is easy to show\cite{ziol} that the field near the focus is given by $H_p(x^{\prime},y^{\prime})$, where the arguments originate at the focal point. This function is not symmetric with respect to $x^{\prime}$ and $y^{\prime}$: the width of 
the distribution in  perpendicular to the slab direction  is larger than in
 the lateral direction (See  Fig.
\ref{fig3}).{\em  This general result originates from the withdrawal of EW's.} Note that
 the source field depends only  on $\rho$. Another important feature of this imaging is that the  image in the lateral direction is  $H_p(0,y^{\prime})=iH_0J_0(y^{\prime} k_0)\exp{-i\omega t}$. If we consider real fields, the part of the source field that is proportional to $\sin(\omega t)$, namely $H_0J_0(\rho k_0)\sin(\omega t)$, has a perfect image in the lateral direction while the  part of the source
with $\cos(\omega t)$ has a zero image in this direction.  This  follows only from the plane (non-axial) geometry of the lens that is assumed to be infinite 
in the lateral direction.  The perfect lateral imaging of the Bessel function $J_0$ does not contradict to general principles because this Bessel function obeys homogeneous Helmholtz equation.

 In our simulation, however,  the slab  has a finite length  $h$ in y-direction. In this case some propagating  modes go outside  the slab. To take this into account one should integrate in Eq. (\ref{p}) from $-k_m$ to $k_m$, where
$k_m=k_0(h/2a)/\sqrt{1+(h/2a)^2}$, $a$ is the distance from the source to the slab. In our case $h/2a=4,8$ for thick and thin lenses respectively. Due to this factor the lateral distribution slightly differs from $J_0^2$.  These calculations are shown in Fig. 3 for both  thick
 (20d)  and thin (10d) lenses. They are different because of different values of $a$. For the thick lens analytical calculation is  in a very good agreement with the computational data. 

 A  sharp first peak of a  Bessel-like  function in lateral direction is  often considered as a result of ``superlensing''\cite{wang,pod}.   We  think, however, that any  result that follows from  Eq.(\ref{p}) has nothing  to do with ``superlensing''.  Following Pendry we would define superlensing as a result of amplification of EW's, but  Eq.(\ref{p})  does not contain EW's at all. It gives just a diffraction pattern of the Veselago lens that
 was  obtained in Ref.\cite{pok3} for a   3D-lens.
 Then the energy distribution  given by this equation should be called 
``regular lensing''. Thus, our     thick Veselago lens based upon PC does not provide any superlensing. The thin lens with AH sufrace gives  substantially  sharper image than it follows from  Eq.(\ref{p}) with a proper $h/2a$-correction(See Fig. 3). This is due to amplification of EW's shown in Fig. 2(c) and 
this effect can be considered as a superlensing.

In conclusion we have shown that the Veselago lens imaging  has some peculiar  features, such as a perfect imaging of the phase shifted part of the source. We show that a thin lens in a near field regime may provide a substantial superlensing depending on the way  the  slab is terminated. The physics of
this superlensing is not connected with the left-handed properties of a medium.

\begin{acknowledgments}
The work has been funded by the NSF grant DMR-0102964 and by the seed grant of the University of Utah. 
\end{acknowledgments}

\bibliography{evan}

\end{document}